\documentclass[aip,amsmath,amssymb,preprint,numerical]{revtex4-1}
\usepackage{graphicx}
\usepackage{epstopdf}
\usepackage{verbatim}
\usepackage{subfigure}
\usepackage{bm}

\begin{document}

\newcommand{\subeff}{ \mathrm{eff} }
\newcommand{\subin}{ \mathrm{in} }
\newcommand{\subex}{ \mathrm{ex} }
\newcommand{\subsq}{ \mathrm{sq} }
\newcommand{\subtri}{ \mathrm{tri} }

\allowdisplaybreaks

\title{Enhanced inertia from lossy effective fluids using multi-scale sonic crystals}

\author{Matthew D. Guild}
\email{mdguild@utexas.edu}
\altaffiliation{Current address: U.S. Naval Research Laboratory, Washington DC 20375, USA. e-mail: mdguild@utexas.edu}
\affiliation{Grupo de Fen\'{o}menos Ondulatorios, Departamento de Ingenieria Electr\'{o}nica,
Universitat Polit\`{e}cnica de Val\`encia, Camino de vera s/n, E-46022 Valencia, Spain}
\author{Victor M. Garcia-Chocano}
\affiliation{Grupo de Fen\'{o}menos Ondulatorios, Departamento de Ingenieria Electr\'{o}nica,
Universitat Polit\`{e}cnica de Val\`encia, Camino de vera s/n, E-46022 Valencia, Spain}
\author{Weiwei Kan}
\affiliation{Grupo de Fen\'{o}menos Ondulatorios, Departamento de Ingenieria Electr\'{o}nica,
Universitat Polit\`{e}cnica de Val\`encia, Camino de vera s/n, E-46022 Valencia, Spain}
\affiliation{Department of Physics, Key Laboratory of Modern Acoustics, MOE, Institute of Acoustics, 
Nanjing University, Nanjing 210093, People's Republic of China}
\author{Jos\'{e} S\'{a}nchez-Dehesa}
\email{jsdehesa@upv.es}
\affiliation{Grupo de Fen\'{o}menos Ondulatorios, Departamento de Ingenieria Electr\'{o}nica,
Universitat Polit\`{e}cnica de Val\`encia, Camino de vera s/n, E-46022 Valencia, Spain}

\date{\today}

\begin{abstract}
In this work, a recent theoretically predicted phenomenon of enhanced permittivity with electromagnetic waves using lossy materials is investigated for the analogous case of mass density and acoustic waves, which represents \emph{inertial enhancement}.  Starting from fundamental relationships for the homogenized quasi-static effective density of a fluid host with fluid inclusions, theoretical expressions are developed for the conditions on the real and imaginary parts of the constitutive fluids to have inertial enhancement, which are verified with numerical simulations.  Realizable structures are designed to demonstrate this phenomenon using multi-scale sonic crystals, which are fabricated using a 3D printer and tested in an acoustic impedance tube, yielding good agreement with the theoretical predictions and demonstrating enhanced inertia.
\end{abstract}

\maketitle

%%%%%%%%%%%%%%%%%%%%%%%%%%%%%%%%%%%%%%%%%
\section{Introduction}

Sonic crystals are periodic lattices of acoustic scatterers which have been utilized as structures for a wide range of acoustic applications, associated with both the acoustic bandgaps at higher frequencies\cite{Dowling1992,Sigalas1992,Kuswaha1993,Sanchezperez1998} in addition to the quasi-static behavior as an effective, homogenized fluid\cite{Krokhin2003,Torrent2006}.  In recent years, the quasi-static nature of sonic crystals has found renewed interest due to enabling the design of acoustic metamaterials, which utilize the dynamics of a microstructural arrangement to produce extreme macroscopic  properties, such as acoustic metafluids with anisotropic inertia\cite{Torrent2008a,Zigoneanu2011}.   Using a homogenized fluid or acoustic metafluid has found great interest in proposed acoustic metamaterial devices, including transformation acoustic cloaking, acoustic scattering cancellation and acoustic hyperlenses \cite{Cummer2008,Torrent2008,Li2009,Sanchis2013,Guild2014}.

Most research on acoustic metamaterials and metafluids have focused on utilization with idealized lossless materials and have sought to minimize the effects of inherent losses in real systems.  Recently, there has been an interest in utilizing acoustic metamaterials for sound absorption applications, which has lead to a more detailed look at the effects of losses in acoustic metamaterials \cite{Sanchez2011,Garcia2012,Christensen2014,Frenzel2013,Climente2012}.  While much of this work has focused on resonant structures such as membranes and mass-spring-damper systems \cite{Naify2010, Yang2010, Hussein2013}, several recent works have investigated the homogenized properties of sonic crystals in viscous fluids \cite{ReyesAyona2012, Guild2014a}, demonstrating that the complex-valued acoustic properties of sonic crystals could be formulated and experimentally verified.

Recently, theoretical and experimental approaches have examined the homogenized properties of lattices with complex-valued permittivities for electromagnetic (EM) waves\cite{Carbonell2010,Carbonell2011,Godin2013}.  Particularly, the theoretical work by Godin\cite{Godin2013} observed that for some cases the homogenized properties of complex-valued constituent materials exhibited a non-monotonic variation with respect to the filling fraction, leading to a maximum value which exceeds the bounds of either constituent material for both the real and imaginary parts of the permittivity.  Expanding this analysis to acoustics can provide insight into the homogenized effective density of lossy acoustic media, and provide a means for enhancing the acoustic properties and expanding the traditional bounds for a composite structure.

In this work, enhancement of homogenized effective properties will be examined for acoustic waves and a formulation of the conditions for inertial enhancement will be presented, which is detailed in Sec.~\ref{Sec:Theory}.  These results are then considered as a function of the filling fraction, which are illustrated for several examples and verified with finite element simulations in Sec.~\ref{Sec:Comsol}.  For realization of the necessary complex-valued effect fluids, the use of multi-scale sonic crystals are proposed, which enable the combination of two arbitrary sonic crystal lattices.  This process is described in Sec.~\ref{Sec:MultiscaleSC}, and the fabrication and experimental testing of multi-scale sonic crystal structures demonstrating enhanced inertia are presented in Sec.~\ref{Sec:Experiment}.

%%%%%%%%%%%%%%%%%%%%%%%%%%%%%%%%%%%%%%%%%
\section{Background} \label{Sec:Background}

To determine the applicability and relevance of Godin's results\cite{Godin2013} to acoustics, one must first examine the relationship between the EM variables and acoustic variables.  The governing equation employed for the electric potential $u$ is \cite{Godin2013}
\begin{equation} \label{Eq:GodinWaveEq}
	\nabla \cdot \left[ \varepsilon(\mathbf{r}) \,\nabla u(\mathbf{r}) \right] = 0,
\end{equation}
\noindent where $\varepsilon(\mathbf{r})$ is the permittivity (which is a function of the spatial position).  
Note that even though Godin defines $u$ and $\varepsilon$ as functions of frequency, Eq.~(\ref{Eq:GodinWaveEq}) is actually Laplace's equation, which is equal to the Helmholtz equation in the limit of $\omega \! \to \! 0$.  
After homogenizing the solution for the electric potential, the resulting effective permittivity is given for an arbitrary periodic lattice arrangement.  For the case of an isotropic lattice based on a periodic distribution of cylinders, these results simplify to give \cite{Godin2013}
\begin{equation} \label{Eq:EpsEff}
	\varepsilon_{\subeff} = \varepsilon_{\subex} \frac{1 +  \frac{\varepsilon_{\subin}  - \varepsilon_{\subex}}{\varepsilon_{\subin}  + \varepsilon_{\subex}} \Lambda f}{1 -  \frac{\varepsilon_{\subin}  - \varepsilon_{\subex}}{\varepsilon_{\subin}  + \varepsilon_{\subex}} \Lambda f},
\end{equation}
where $\varepsilon_{\subin}$ and $\varepsilon_{\subex}$ correspond to the permittivity in interior (cylinders) and exterior (surrounding medium), respectively, and $\Lambda$ is a geometry-dependent coefficient obtained from the expansion of the electric potential.

%%% Figure 1
\begin{figure}[t!]
	\includegraphics[width=0.99\columnwidth, height=0.7\textheight, keepaspectratio]{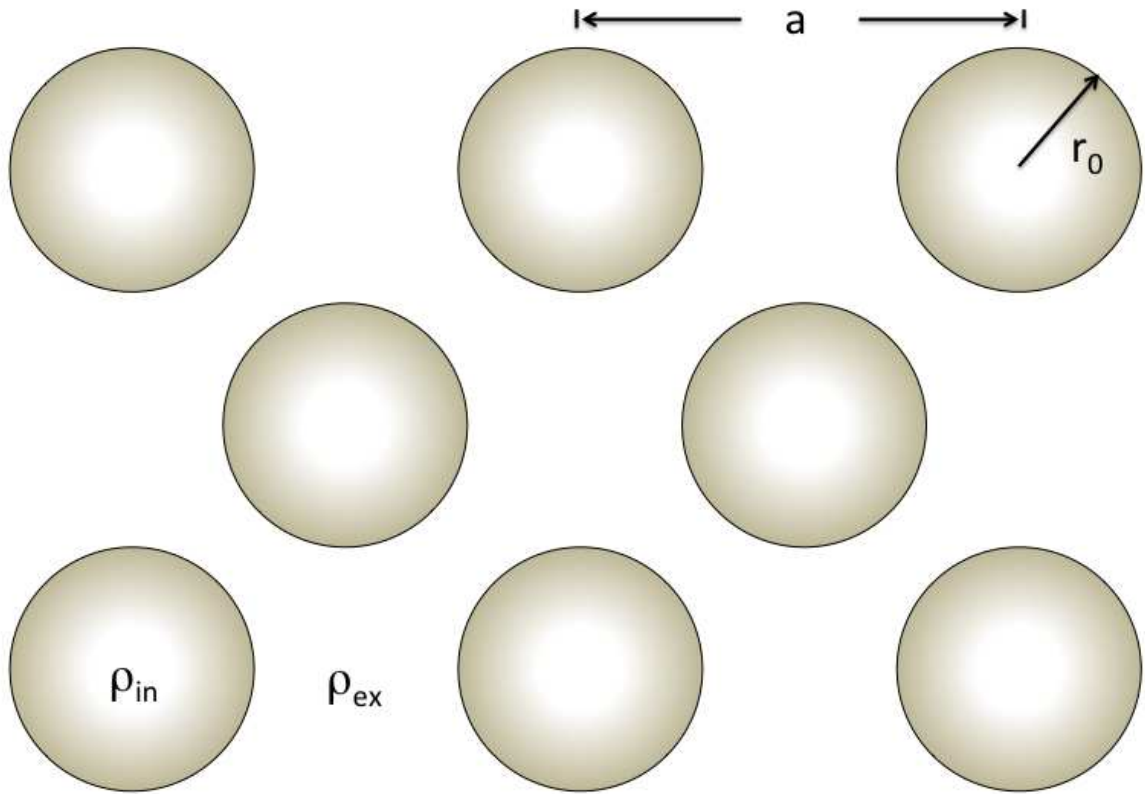}
	\caption{Geometry for a sonic crystal with lattice parameter $a$ and cylinder radius $r_{0}$.  The complex-valued density of the host (exterior) medium is $\rho_{\subex}$ and $\rho_{\subin}$ for the inclusion (interior) medium.}
	\label{Fig:LatticeGeom}
\end{figure}

The equivalent behavior in an acoustic system can be examined by considering the effective fluid density.  In the quasi-static limit, the effective density for an infinite lattice of fluid cylinders has been extensively studied, and can be written as \cite{Torrent2006}
\begin{equation} \label{Eq:RhoEff1}
	\rho_{\subeff} = \rho_{\subex} \frac{\rho_{\subin}(\Delta + f) + \rho_{\subex}(\Delta - f)}{\rho_{\subin}(\Delta - f) + \rho_{\subex}(\Delta + f)},
\end{equation}
\noindent where $\rho_{\subex}$ is the fluid density of the exterior fluid and $\rho_{\subin}$ is the fluid density of the cylinders as illustrated in Fig.~\ref{Fig:LatticeGeom}, and $\Delta$ is a coefficient which depends on the geometry and filling fraction of the lattice.  The coefficient $\Delta$ has previously been written as an expansion to include a leading order term proportional to $f^{2}$, but for low to moderate filling fractions it has been shown that \cite{Torrent2007} $\Delta \! \approx \! 1$.  Equation~(\ref{Eq:RhoEff1}) can be rewritten in a form similar to that of Eq.~(\ref{Eq:EpsEff}), such that
\begin{equation} \label{Eq:RhoEff2}
	\rho_{\subeff} = \rho_{\subex} \frac{1 + \frac{1}{\Delta} \left[ \frac{\rho_{\subin} \,-\, \rho_{\subex}}{\rho_{\subin} \,+\, \rho_{\subex}} \right] f}{1 - \frac{1}{\Delta} \left[ \frac{\rho_{\subin} \,-\, \rho_{\subex}}{\rho_{\subin} \,+\, \rho_{\subex}} \right] f}.
\end{equation}
This is identical to the expression for $\rho_{\subeff}$ by the acoustic analogy to the solution developed for electromagnetic waves, except that the $\Lambda$ term has been replaced by $1/\Delta$.

While an exact formulation for $\Delta$ has been given previously \cite{Torrent2007}, an approximate form is sought in terms of the filling fraction and neglecting higher order (multiple interaction) scattering terms.  The effective density accounting for single scattering effects can be expanded in terms of the filling fraction, which gives \cite{Martin2010}
\begin{equation} \label{Eq:RhoEffMartin}
	\rho_{\subeff} \approx \rho_{\subex} \left[ 1 + j\frac{8}{\pi} \frac{Z_{1}}{\left( k_{\subex}r_{0} \right)^{2}} f - \frac{32}{\pi^{2}} \frac{Z_{1}^{2}}{\left( k_{\subex}r_{0} \right)^{4}} f^{2} \right],
\end{equation}
\noindent where
\begin{equation} \label{Eq:Zcoef}
	Z_{n} = \frac{ \frac{\rho_{\subin}}{\rho_{\subex}} J_{n}'(k_{\subex}r_{0}) J_{n}(k_{\subin}r_{0}) - \frac{k_{\subin}}{k_{\subex}} J_{n}'(k_{\subin}r_{0}) J_{n}(k_{\subex}r_{0})  }{  \frac{\rho_{\subin}}{\rho_{\subex}} H_{n}'(k_{\subex}r_{0}) J_{n}(k_{\subin}r_{0}) \!-\! \frac{k_{\subin}}{k_{\subex}} J_{n}'(k_{\subin}r_{0}) H_{n}(k_{\subex}r_{0})  },
\end{equation}
\noindent with $J_{n}$ and $H_{n}$ denoting the Bessel and Hankel functions, and the prime denoting the first derivative.  Expanding Eq.~(\ref{Eq:RhoEff2}) and matching terms, one finds that
\begin{equation} \label{Eq:RhoEffApprox}
	\rho_{\subeff} \approx \rho_{\subex} \left\{ 1 + \frac{2}{\Delta}\left[ \frac{\rho_{\subin} \,-\, \rho_{\subex}}{\rho_{\subin} \,+\, \rho_{\subex}} \right] f + \frac{2}{\Delta^{2}}\left[ \frac{\rho_{\subin} \,-\, \rho_{\subex}}{\rho_{\subin} \,+\, \rho_{\subex}} \right]^{2} f^{2} \right\},
\end{equation}
\noindent which by comparison with Eq.~(\ref{Eq:RhoEffMartin}) gives
\begin{equation} \label{Eq:DeltaApprox}
	\Delta \approx -j \frac{\pi}{4}  \frac{\left( k_{\subex}r_{0} \right)^{2}}{Z_{1}} \left[ \frac{\rho_{\subin} \,-\, \rho_{\subex}}{\rho_{\subin} \,+\, \rho_{\subex}} \right].
\end{equation}
Note that when $k_{\subex}r_{0} \! \ll \! 1$, Eq.~(\ref{Eq:Zcoef}) simplifies to give \cite{Martin2010}
\begin{equation} \label{Eq:ZcoefSmallArg}
	Z_{1} \approx -j \frac{\pi}{4} \left( k_{\subex}r_{0} \right)^{2} \left[ \frac{\rho_{\subin} \,-\, \rho_{\subex}}{\rho_{\subin} \,+\, \rho_{\subex}} \right],
\end{equation}
\noindent and therefore $\Delta \! \approx \! 1$ as expected.  Thus, although  Eq.~(\ref{Eq:RhoEff2}) is relatively simple and strictly valid only in the quasi-static limit, it can be applied to more moderate frequencies and filling fractions of interest where single scattering effects become important through the use of the coefficient $\Delta$ in its approximate form given by Eq.~(\ref{Eq:DeltaApprox}).

%%%%%%%%%%%%%%%%%%%%%%%%%%%%%%%%%%%%%%%%
\section{Theoretical formulation for enhanced inertia} \label{Sec:Theory}

To examine the conditions necessary to achieve an enhancement of the complex inertia, the complex densities of the effective lossy fluids in the exterior and interior media can be expressed as
\begin{align}
	\rho_{\subex} &= \rho_{0} \!-\! j \rho_{0}', \label{Eq:RhoEx} \\
	\rho_{\subin} &= \rho - j \rho'.  \label{Eq:RhoIn} 
\end{align}
Use of these expressions in Eq.~(\ref{Eq:RhoEff1}) enables the effective density to be written in terms of the real and imaginary terms, which yields
\begin{equation}  \label{Eq:RhoEff_RealImag}
	\rho_{\subeff} = \left[  \frac{\alpha \rho_{0} -  \beta \rho_{0}'}{\gamma} \right] - j \left[ \frac{ \beta\rho_{0} + \alpha \rho_{0}'}{\gamma} \right],
\end{equation}
\noindent where
\begin{align}
	\alpha &= [\rho (\Delta \!+\! f) \!+\! \rho_{0}(\Delta \!-\! f)] [\rho (\Delta \!-\! f) \!+\! \rho_{0} (\Delta \!+\! f) ] \notag \\
		&+ [\rho' (\Delta \!+\! f) \!+\! \rho_{0}'(\Delta \!-\! f)] [\rho' (\Delta \!-\! f) \!+\! \rho_{0}' (\Delta + f) ] ,\label{Eq:RhoEff_Alpha} \\
	\beta &=  4 f \Delta [ \rho_{0} \rho' - \rho \rho_{0}' ] ,  \label{Eq:RhoEff_Beta} \\
	\gamma &= [\rho (\Delta \!-\! f) \!+\! \rho_{0} (\Delta \!+\! f) ]^{2} \!+\! [\rho' (\Delta \!-\! f) \!+\! \rho_{0}' (\Delta + f) ]^{2}.   \label{Eq:RhoEff_gamma} 
\end{align}

For the \emph{enhancement} of the effective density, this corresponds to two different possibilities, where either the \emph{real} or \emph{imaginary} part are greater than the maximum value of either of the constituent materials.  Thus, it follows from Eq.~(\ref{Eq:RhoEff_RealImag}) that the criteria for these two cases can be expressed as
\begin{align}
	\left[  \frac{\alpha \rho_{0} -  \beta \rho_{0}'}{\gamma} \right] &> \rho_{\mathrm{max}}, \label{Eq:EnhancedInertia_Real} \\
	\left[ \frac{ \beta\rho_{0} + \alpha \rho_{0}'}{\gamma} \right] &> \rho_{\mathrm{max}}', \label{Eq:EnhancedInertia_Imag}
\end{align}
\noindent for \emph{real} and \emph{imaginary} inertia enhancement, respectively, where $\rho_{\mathrm{max}} \!=\! \mathrm{max}(\rho,\rho_{0})$ and $\rho_{\mathrm{max}}' \!=\! \mathrm{max}(\rho',\rho_{0}')$.  To examine Eqs.~(\ref{Eq:EnhancedInertia_Real}) and (\ref{Eq:EnhancedInertia_Imag}) further, we will now consider the case where the imaginary part in each medium is much less than that of the real part, in which case the relationships between $\alpha$, $\beta$ and $\gamma$ simplify to yield
\begin{align}
	\frac{\beta}{\alpha} &\approx  \frac{4 f \Delta [ \rho_{0} \rho' - \rho \rho_{0}' ] }{[\rho (\Delta \!+\! f) \!+\! \rho_{0}(\Delta \!-\! f)] [\rho (\Delta \!-\! f) \!+\! \rho_{0} (\Delta \!+\! f) ]},  \label{Eq:Beta2Alpha} \\
	\frac{\gamma}{\alpha} &\approx \frac{[\rho (\Delta \!+\! f) \!+\! \rho_{0}(\Delta \!-\! f)]}{[\rho (\Delta \!-\! f) \!+\! \rho_{0} (\Delta \!+\! f) ]} = \frac{\rho_{0}}{\rho_{\mathrm{eff}}^{(0)} },   \label{Eq:Gamma2Alpha} 
\end{align}
\noindent where $\rho_{\mathrm{eff}}^{(0)}$ is the effective density given by Eq.~(\ref{Eq:RhoEff1}) without losses, which corresponds to when $\rho_{\mathrm{ex}}$ and $\rho_{\mathrm{in}}$ are real.

%%%%%%%%%%%%%%%%%%%
\subsection{Enhancement of $\mathrm{Im} \! \left[ \rho_{\mathrm{eff}} \right]$ } \label{Sec:ImRhoEff}
The first case is that of enhancement of the imaginary part of the inertia, which is prescribed by Eq.~(\ref{Eq:EnhancedInertia_Imag}).  For ordinary composites the imaginary part would be increased by using inclusions with $\rho' \!>\! \rho_{0}'$, resulting in an imaginary part of the effective density greater that $\rho_{0}'$ but less than $\rho'$.  Alternatively, we will consider the enhanced imaginary inertia for the case where $\rho' \!<\! \rho_{0}'$, which for an ordinary composite would lead to a \emph{decrease} in the imaginary effective density.  In this case, $\rho_{\mathrm{max}}' \!=\! \rho_{0}'$, and substitution of Eq.~(\ref{Eq:Beta2Alpha} ) and (\ref{Eq:Gamma2Alpha} ) into Eq.~(\ref{Eq:EnhancedInertia_Imag}) yields the following criteria on $\rho$ for enhancement:
\begin{equation}  \label{Eq:Imag_RhoQuad}
	\rho^{2} - 2 \rho + \rho_{0} \left[ \frac{2 \Delta \rho' - \rho_{0}' ( \Delta + f) }{\rho_{0}' (\Delta \!-\! f)} \right] > 0.
\end{equation}
To determine the critical value of $\rho$ at which the enhancement occurs, we consider the case where the lefthand side of Eq.~(\ref{Eq:Imag_RhoQuad}) is identically equal to zero, so that  $\rho \!>\! \rho_{\mathrm{crit}}$, where
\begin{equation}  \label{Eq:Imag_RhoQuad}
	\rho_{\mathrm{crit}}^{2} - 2 \rho_{\mathrm{crit}} + \rho_{0} \left[ \frac{2 \Delta \rho' - \rho_{0}' ( \Delta + f) }{\rho_{0}' (\Delta \!-\! f)} \right] = 0.
\end{equation}
\noindent Equation~(\ref{Eq:Imag_RhoQuad}) is simply a quadratic equation in terms of $\rho_{\mathrm{crit}}$, which can be solved to obtain
\begin{equation}  \label{Eq:Imag_RhoQuadCrit}
	\rho_{\mathrm{crit}} =  \rho_{0} \left[ 1+ \sqrt{ \frac{2}{ \left(1 \!-\! \frac{f}{\Delta} \right)} \left[ 1 \!-\! \frac{\rho'}{\rho_{0}'} \right] } \right],
\end{equation}
\noindent Given that $\rho' \!<\! \rho_{0}'$, the square root term in Eq.~(\ref{Eq:Imag_RhoQuadCrit})  will be positive, ensuring a real solution (though leading to an increasingly large value of $\rho_{\mathrm{crit}}$ for $f/\Delta \!\to\! 1$).  Since $\rho \!>\! \rho_{\mathrm{crit}}$, this gives the condition for enhancement of $\mathrm{Im} \! \left[ \rho_{\mathrm{eff}} \right]$:
\begin{equation}  \label{Eq:Imag_RhoCrit}
	\rho >  \rho_{0} \left[ 1+ \sqrt{ \frac{2}{ \left(1 \!-\! \frac{f}{\Delta} \right)} \left[ 1 \!-\! \frac{\rho'}{\rho_{0}'} \right] } \right], \quad \rho' < \rho_{0}'.
\end{equation}
Thus, it is apparent that the real part of the interior (inclusion) density must be larger than the real part of the exterior (host) density.  In fact, this highlights the underlying physical mechanism by which the enhancement occurs: that it is the large values of the \emph{real} parts of the density which are leading to an increase in the homogenized \emph{imaginary} part.  In particular, this coupling between the real and imaginary parts can be traced back to the cross products that result during the process of separating $\rho_{\mathrm{eff}}$ into real and imaginary terms in Eq.~(\ref{Eq:RhoEff_RealImag}).

%%%%%%%%%%%%%%%%%%%
\subsection{Enhancement of $\mathrm{Re} \! \left[ \rho_{\mathrm{eff}} \right]$ } \label{Sec:ReRhoEff}
Similar to the result of enhancement of $\mathrm{Im} \! \left[ \rho_{\mathrm{eff}} \right]$ which arose from an increase due to \emph{real} components of the density, one would be expected that a correspondingly large increase in the imaginary part could result in the enhancement of $\mathrm{Re} \! \left[ \rho_{\mathrm{eff}} \right]$.  To examine this, let us consider the condition prescribed by Eq.~(\ref{Eq:EnhancedInertia_Real}) for the enhancement of the real part of the effective density, which can be rewritten through the use of Eqs.~(\ref{Eq:Beta2Alpha}) and (\ref{Eq:Gamma2Alpha}) to obtain
\begin{align}
\delta_{\mathrm{lossless}} &-  \frac{4 f \Delta [ \rho_{0} \rho' - \rho \rho_{0}' ] \frac{ \rho_{0}'}{ \rho_{0} } \frac{\rho_{\mathrm{eff}}^{(0)} }{ \rho_{\mathrm{max}} } }{[\rho (\Delta \!+\! f) \!+\! \rho_{0}(\Delta \!-\! f)] [\rho (\Delta \!-\! f) \!+\! \rho_{0} (\Delta \!+\! f) ]} > 0, \label{Eq:EnhancedInertia_RealCrit1} \\
\delta_{\mathrm{lossless}} &= \left[\frac{\rho_{\mathrm{eff}}^{(0)} }{ \rho_{\mathrm{max}} } - 1 \right]. \label{Eq:EnhancedInertia_RealCrit2}
\end{align}
From Eq.~(\ref{Eq:EnhancedInertia_RealCrit2}), it can be observed that $\delta_{\mathrm{lossless}}$ is associated with the contribution from the lossless component of the effective density, $\rho_{\mathrm{eff}}^{(0)}$.  However, $\rho_{\mathrm{eff}}^{(0)} \! \le \! \rho_{\mathrm{max}} $, which means $\delta_{\mathrm{lossless}} \! \le \! 0$, and thus this term \emph{inhibits} the enhancement prescribed by Eq.~(\ref{Eq:EnhancedInertia_RealCrit1}).  To obtain the optimal conditions for enhancement we seek the condition where $\delta_{\mathrm{lossless}} \!=\! 0$ (and therefore $\rho_{\mathrm{eff}}^{(0)} \! = \! \rho_{\mathrm{max}}$), which occurs when $\rho \! = \! \rho_{0}$.  In this case Eq.~(\ref{Eq:EnhancedInertia_Real}) reduces to
\begin{equation}  \label{Eq:Real_RhoUnityCond1}
	\frac{\rho_{0}'}{\rho_{0}} > \frac{\rho'}{\rho},
\end{equation}
\noindent or, since $\rho \! = \! \rho_{0}$, this expression is simply given by $\rho_{0}' \! > \! \rho'$.  This result shows that to obtain the enhancement of $\mathrm{Re} \! \left[ \rho_{\mathrm{eff}} \right]$, one should seek the same real part of the density in both the exterior and interior materials, with the exterior material having a larger imaginary part than the interior. Analogous to the results observed in the previous section, it is the \emph{imaginary} parts of the density which describe the conditions for enhancement of the \emph{real} part of the effective density.

%%%%%%%%%%%%%%%%%%%%%%%%%%%%%%%%%%%%%%%%%
\section{Comparison of results with finite element simulations} \label{Sec:Comsol}

To highlight and verify the interesting features of the enhanced homogenized inertia given in Sec.~\ref{Sec:ImRhoEff} and \ref{Sec:ReRhoEff}, these results will be compared with simulations based on the finite element method.  The calculations were performed using the commercially available software Comsol Multiphysics. A 2D simulation of an acoustic domain containing circular fluid inclusions imbedded in an exterior fluid medium with complex-valued parameters was utilized to verify the results.  By creating a finite slab of this composite structure, the effective acoustic properties were extracted using well-established techniques \cite{Song2000,Guild2014a}.  Each point in the data set was obtained by varying the radius and spacing of the lattice to achieve the desired filling fraction.

\begin{figure}[t!]
	\includegraphics[width=0.99\columnwidth, height=0.7\textheight, keepaspectratio]{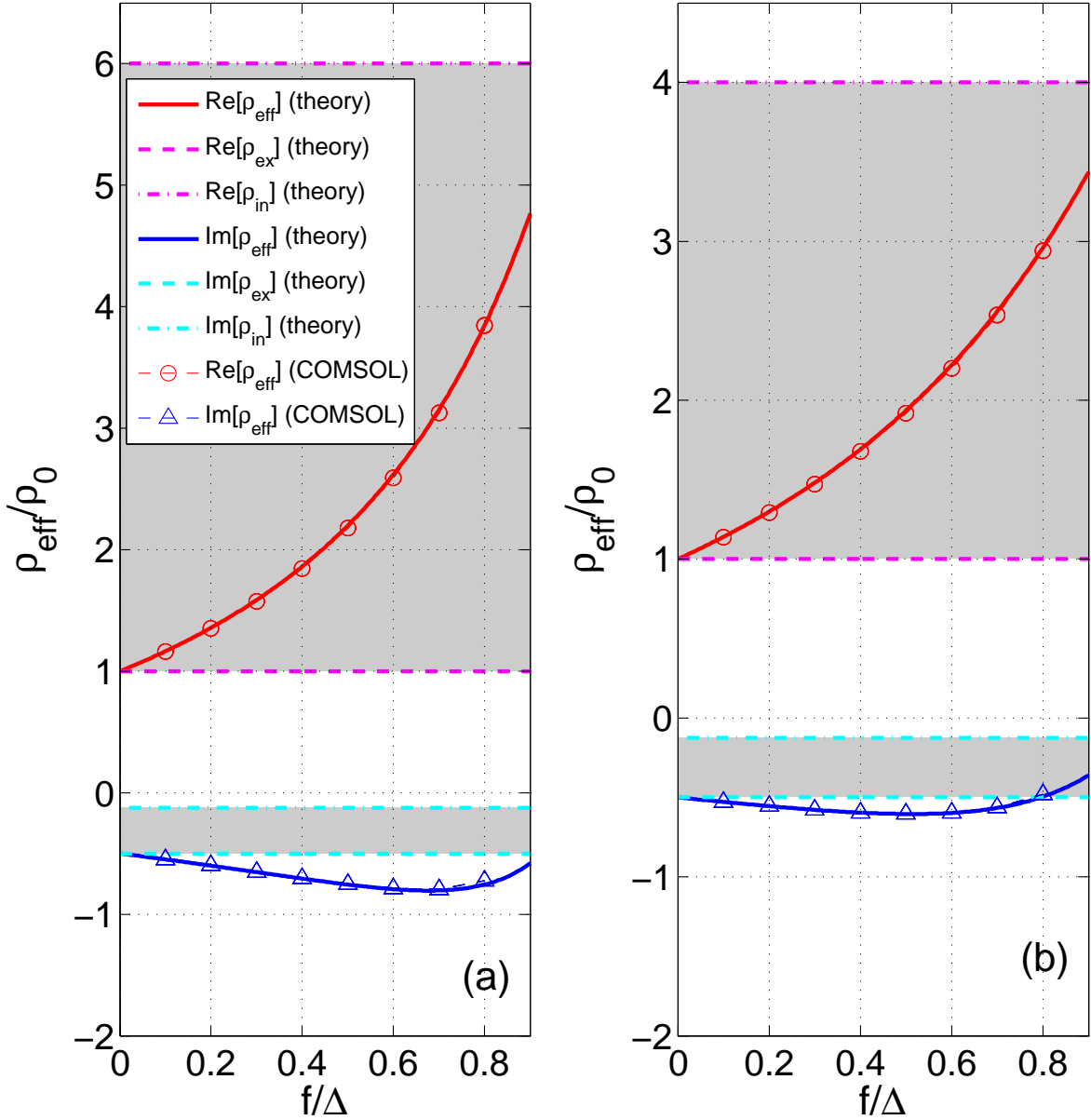}
	\caption{Comparison of analytical and numerical (Comsol) results as a function of reduced filling fraction for $\rho_{\mathrm{ex}} \!=\! 1 \!-\! 0.5j$, with (a) $\rho_{\mathrm{in}} \!=\! 6 \!-\! 0.125j$ and (b) $\rho_{\mathrm{in}} \!=\! 4 \!-\! 0.125j$ }
	\label{Fig:EnhancedInertia_Imag}
\end{figure}

Two specific cases are presented, which utilize complex-valued properties for both the interior and exterior fluids to create enhanced inertia, as predicted by the theoretical formulation developed in the previous sections.  The first case to be considered is selected to demonstrate enhancement of the imaginary part of the effective density, which is illustrated in Figure~\ref{Fig:EnhancedInertia_Imag} as a function of the reduced filling fraction, $f/\Delta$.  For this case, the exterior fluid has a complex density of $\rho_{\mathrm{ex}} \!=\! 1 \!-\! 0.5j$, with two different fluid inclusions $\rho_{\mathrm{in}} \!=\! 6 \!-\! 0.125j$ and $\rho_{\mathrm{in}} \!=\! 4 \!-\! 0.125j$, presented in Figure~\ref{Fig:EnhancedInertia_Imag}(a) and (b), respectively. As shown, the resulting real part is observed to vary monotonically between the bounds of the exterior and interior values (denoted by the dashed and dash-dotted lines, respectively), whereas the magnitude of the imaginary part \emph{exceeds} the bounds of ordinary composites, which are denoted by the shaded regions between the respective densities of the constituent components.  There is excellent agreement between the analytic results obtained with Eq.~(\ref{Eq:RhoEff2}) and Comsol simulations for both the real and imaginary parts and across the entire range of filling fractions examined. Furthermore, the observed enhancement is identical to that numerical investigated previously for the effective permittivity \cite{Godin2013}.

%%% Figure 2

From a physical perspective, the imaginary part of a material property corresponds to the losses in the system. Therefore, the observed increase in the magnitude of the imaginary part means that the effective properties of the homogenized structure will display higher losses than the losses associated with the individual components.  While conceptually very intriguing, such an effect is not uncommon for porous acoustic absorbers, which are made up of a stiff porous structure (often modeled as a rigid, lossless frame) with a viscous fluid filling the pores.  The increased viscous friction between the fluid and the pore walls leads to significantly higher losses than either the frame or the viscous fluid alone.  However, it is interesting that in the results presented in Fig.~\ref{Fig:EnhancedInertia_Imag}, no physical mechanism for losses are included in the analysis, such as thermal or viscous effects for the acoustic system, and therefore the enhanced losses observed are due solely to the quasi-static homogenization of complex-valued (lossy) materials.

%%% Figure 3
\begin{figure}[t!]
	\includegraphics[width=0.99\columnwidth, height=0.7\textheight, keepaspectratio]{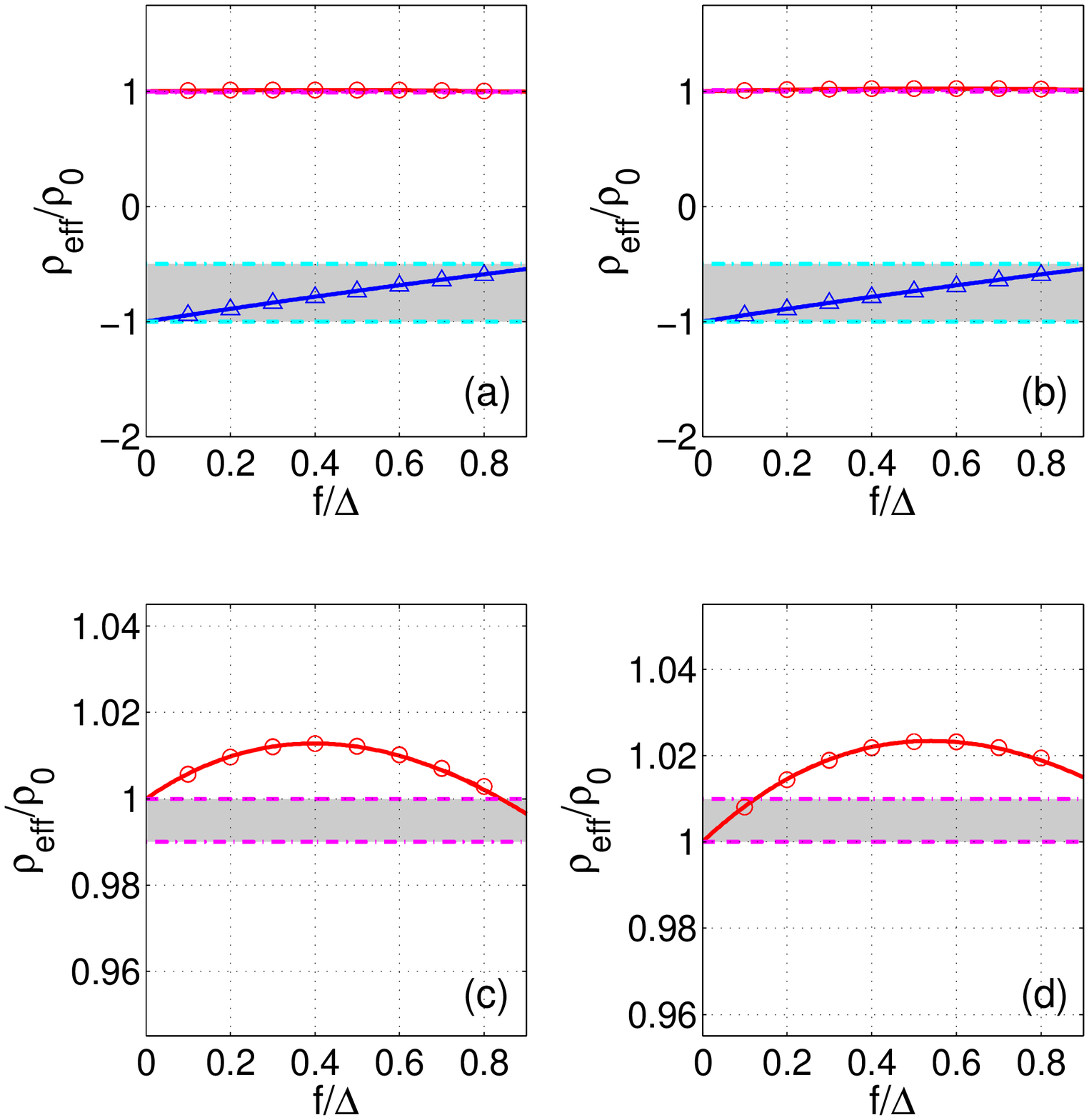}
	\caption{Comparison of analytical and numerical (Comsol) results as a function of reduced filling fraction for $\rho_{\mathrm{ex}} \!=\! 1 \!-\! j$, with (a) $\rho_{\mathrm{in}} \!=\! 0.99 \!-\! 0.5j$ and (b) $\rho_{\mathrm{in}} \!=\! 1.01 \!-\! 0.5j$.  A zoomed-in view of the real part is illustrated for (c) $\rho_{\mathrm{in}} \!=\! 0.99 \!-\! 0.5j$ and (d) $\rho_{\mathrm{in}} \!=\! 1.01 \!-\! 0.5j$. The legend is the same is in Figure~\ref{Fig:EnhancedInertia_Imag}.}
	\label{Fig:EnhancedInertia_Real}
\end{figure}

The results for the second case to be considered demonstrates enhancement of the real part of the effective density, which is illustrated in Fig.~\ref{Fig:EnhancedInertia_Real} as a function of the reduced filling fraction, $f/\Delta$.  For this example, the exterior fluid has a complex density of $\rho_{\mathrm{ex}} \!=\! 1 \!-\! j$, with two different cases of fluid inclusions $\rho_{\mathrm{in}} \!=\! 0.99 \!-\! 0.5j$ and $\rho_{\mathrm{in}} \!=\! 1.01 \!-\! 0.5j$, which are presented in Fig.~\ref{Fig:EnhancedInertia_Real}(a) and (b), respectively.  The resulting imaginary part varies monotonically between the bounds of the exterior and interior values, and it is the real part that exceeds the bounds of either of the individual components.  Due to the small scale of the enhancement relative to the magnitude of the densities of the constitutive components, a zoomed-in view of the real part is presented in Fig.~\ref{Fig:EnhancedInertia_Real}(c) and (d).  For each configuration and filling fraction, excellent agreement is observed between the analytic results and Comsol simulations.

Physically, the observed increase in the real part means that the effective properties of the homogenized structure will display higher properties with respect to the propagating wave than those associated with the individual components.  Furthermore, although the effective quasi-static density of fluid-saturated structures is known to increase the density over static values as described by Eq.~(\ref{Eq:RhoEff1}), this value includes these effects, and therefore the enhancement exceeds even these bounds when compared with those of the lossless case.  A similar enhancement of the real part was also observed for the permittivity with EM waves \cite{Godin2013}.  Although the enhancement illustrated in Fig.~\ref{Fig:EnhancedInertia_Real} is somewhat modest, these values selected represent a realistic range of physically attainable fluid properties, and the realization of such materials and experimental demonstration of enhanced inertia is discussed in the next section.

%%%%%%%%%%%%%%%%%%%%%%%%%%%%%%%%%%%%%%%%%
\section{Realization using multi-scale sonic crystals} \label{Sec:MultiscaleSC}

The theoretical framework for enhanced inertia was presented in the previous sections, for which arbitrary fluids with complex densities were used to create the desired enhancement.  Although theoretically simple, realization of structures made of fluids having specific complex values of density presents significant challenges.  The primary obstacles to realizing this inertial enhancement are associated with creating the necessary complex-valued fluids (particularly those with large imaginary components), and then being able to combine fluid elements with different properties in a precise, structured manner.  It is widely known that complex-valued densities can be used to represent the losses present in real fluids.  However, these losses often arise from viscous and thermal effects which are intrinsic to the fluid, and are difficult to modify to achieve the necessary conditions for inertial enhancement.

%%% Figure 4
\begin{figure}[t!]
	\includegraphics[width=0.99\columnwidth, height=0.5\textheight, keepaspectratio]{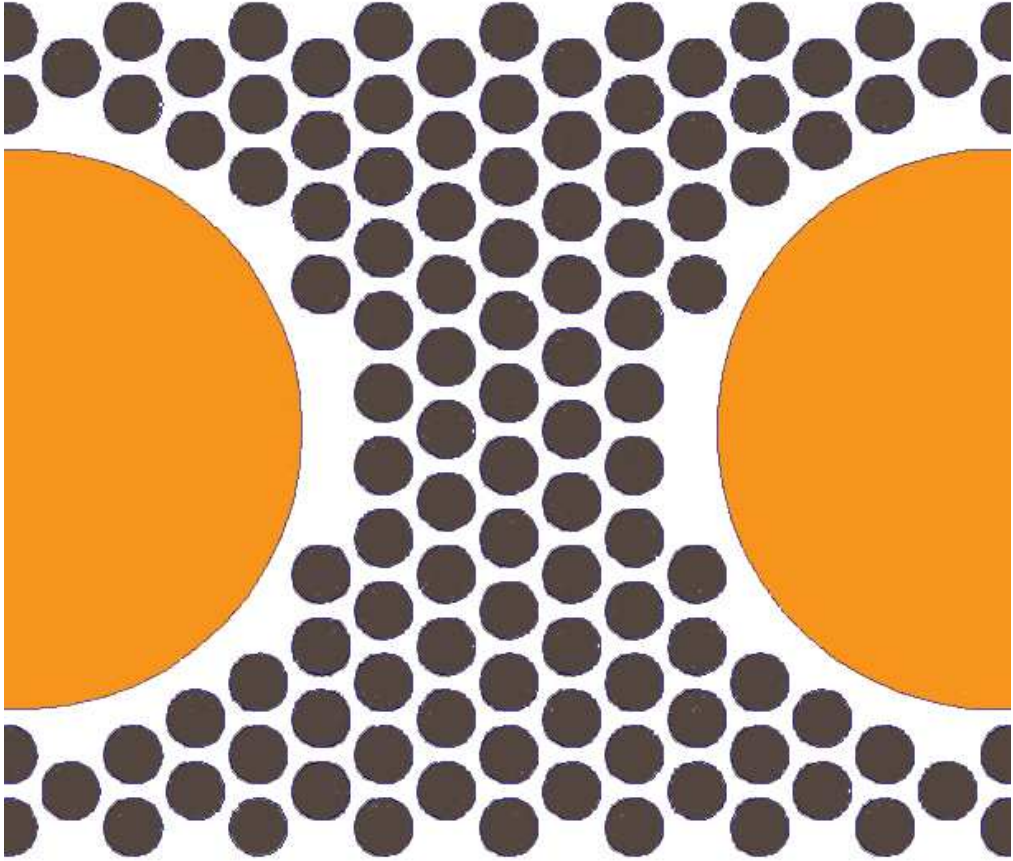}
	\caption{Lattice geometry for the multi-scale sonic crystal examined in this work to demonstrate the enhancement of the imaginary component of the effective density. }
	\label{Fig:MultiscaleGeom}
\end{figure}

One means of overcoming this challenge is through the use of effective fluids created using sonic crystals.  Previously, sonic crystals have been utilized to achieve the acoustic properties of Argon gas using wooden cylinders in air \cite{Torrent2006}.  More recently, this work has been expanded to creating lossy effective fluids, both theoretically and experimentally \cite{Guild2014a, ReyesAyona2012}.  With a plastic lattice made using a 3D printer, the complex-valued effective density and bulk modulus were measured for sonic crystal lattices with moderate filling fractions by the authors, and shown to be in excellent agreement with theoretical results \cite{Guild2014a}.  To accurately account for the imaginary part of the density, the thermoviscous effects must be considered, which for a sonic crystal in air can be expressed as\cite{Guild2014a}
\begin{align}
	\rho_{\mathrm{eff}} &= \rho_{0} \left( \frac{1 \!+\! f}{1 \!-\! f} \right)\left[ 1 - j \frac{\bar{F}_{\mathrm{sc}}}{\bar{\omega}_{\mathrm{sc}}}\right], \label{Eq:Rho_SC} \displaybreak[0] \\
	\bar{F}_{\mathrm{sc}} &= \sqrt{ 1 \!+\! j \frac{1}{2}\bar{\omega}_{\mathrm{sc}} M_{\mathrm{sc}} }, \label{Eq:Fbar_SC} \displaybreak[0] \\
	\bar{\omega}_{\mathrm{sc}} &= \frac{1}{2 f (\frac{\delta}{r_{0}})^{2}} \left( \frac{1 \!+\! f}{1 \!-\! f} \right) \left[ -\frac{1}{2} \ln f - \frac{3}{4} + f -\frac{1}{4}f^{2} \right], \label{Eq:Omega_SC} \displaybreak[0] \\
	M_{\mathrm{sc}} &= \frac{8 f}{\left(1 \!-\! f^{2}\right)^{2}} \! \left( \frac{1 \!+\! f}{1 \!-\! f} \right) \! \left[ -\frac{1}{2} \ln f - \frac{3}{4} + f -\frac{1}{4}f^{2} \right], \label{Eq:M_SC}
\end{align}
\noindent where $\rho_{0}$ is the density of air and $\delta$ is the viscous boundary layer thickness.  Through the use of sonic crystals, an effective medium can be created that acoustically acts as a lossy fluid, but which is mechanically rigid.  This unique combination of characteristics allows for precise geometries and lattices to be realized.  In the present work, sonic crystals are used to obtain fluids with different complex-valued densities to create an effective fluid host (exterior medium) and fluid inclusions (interior medium).  Therefore, Eq.~(\ref{Eq:RhoEff2}) can be applied to determine the density for each lossy effective medium, whose internal $\rho_{\subin}$ and external $\rho_{\subex}$ densities can be calculated from Eq.~(\ref{Eq:Rho_SC})--(\ref{Eq:M_SC}).

Given a means for obtaining effective lossy fluids, the next challenge to be addressed is combining two different sonic crystal structures.  Previous work\cite{Guild2014a} has shown that an acoustic metamaterial absorber made from alternating layers of uniform cylinders (with different lattice parameters in each layer) can be achieved.  However, these results were obtained with sonic crystals possessing uniform cylinders but significantly different filling fractions, which created sufficient contrast between the two effective media.  In general, though, as the contrast between the two different sonic crystal decreases, the two layers will appear more like a single effective fluid.  Alternatively, when the cylinders of the lattice in each effective fluid are significantly different in size, the smaller cylinders will appear as a homogenized medium even for similar filling fractions due to the difference in scale, thereby creating a \emph{multi-scale} structure.

A sample was designed and fabricated using a multi-scale sonic crystal structure. It is intended to demonstrate the enhanced inertia effect for the imaginary part, whose effect is more pronounced than that observed in the real part and lead to more interesting applications since the absorptive properties of the metamaterial are enhanced. The scheme of the sample is illustrated in Fig.~\ref{Fig:MultiscaleGeom}. As shown, the external medium is implemented through a regular lattice of small cylinders. To enhance the imaginary part, an interior medium density $\rho_{\subin}$ with large real and small imaginary part is needed. This condition is satisfied with a sonic crystal having large solid cylinders with a moderate separation between them. It is worth to note that the dimensions of the cylinders defining the external medium are limited by the ability of the 3D printer, while the low number of cylinders of the interior medium is due to the dimensions of the impedance tube where the experiments were carried out.

%%%%%%%%%%%%%%%%%%%%%%%%%%%%%%%%%%%%%%%%%
\section{Experimental results} \label{Sec:Experiment}

%%% Figure 5
\begin{figure}[t!]
	\includegraphics[width=0.99\columnwidth, height=0.5\textheight, keepaspectratio]{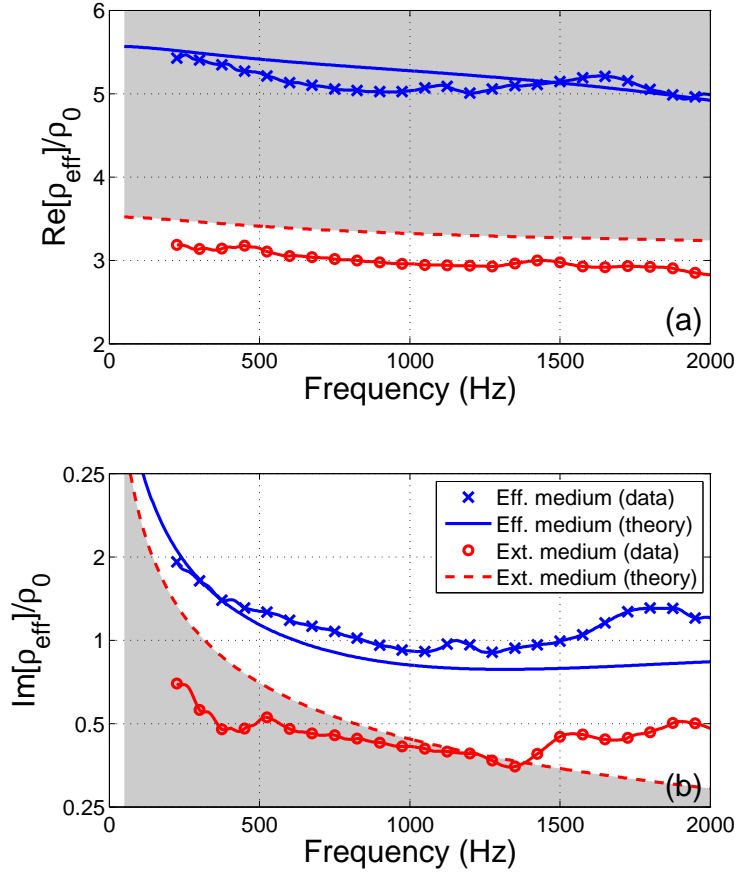}
	\caption{Comparison of theoretical and experimental data for lattice geometry described in Fig.~\ref{Fig:MultiscaleGeom}.}
	\label{Fig:DataCase01}
\end{figure}

To verify the theoretical model for the multi-scale sonic crystals, the sample illustrated in Fig.~\ref{Fig:MultiscaleGeom} was fabricated using ABS plastic by using a 3D printer. It was experimentally tested in a standard circular cross-section acoustic impedance tube, using the same method and equipment described in previous works \cite{Guild2014a}.  In brief, the pressure field inside the tube is measured at four different points (two on each side of the sample) and the effective parameters are obtained through the transfer-matrix formulation introduced in Ref. \onlinecite{Song2000}. The radius of the cylinders are $ r_{\subex}=1mm$ and $ r_{\subin}=9.7mm$ for the external and internal medium, respectively. The lattice constants of the two sonic crystals are $ a_{\subex}=2.5mm$ and $ a_{\subin}=33.7mm$. In addition to the sample, a uniform (single-scale) sonic crystal was constructed with a lattice geometry corresponding to the exterior medium to verify the constituent effective fluid.

A comparison of theoretical and experimental results is presented in Fig.~\ref{Fig:DataCase01}.  In Fig.~\ref{Fig:DataCase01}(a), the \emph{real} part of the effective density normalized by the density of air is presented, with the lines representing theoretical values and the markers representing the experimental data.  The shaded region indicates the ordinary range of a composite structure for this case.  As observed in Fig.~\ref{Fig:DataCase01}(a), the experimental results show that for enhancement of $\mathrm{Im}[\rho_{\subeff}]$, the real part of $\rho_{\subeff}$ lies within the bounds of an ordinary composite, as predicted by the theory.  In Fig.~\ref{Fig:DataCase01}(b), the \emph{imaginary} part of the effective density normalized by the real part of $\rho_{\subex}$ is presented.  There is good agreement between the experimental data and theoretical results, which clearly show an enhancement in the imaginary part of the effective density, nearly twice that of the upper bound for an ordinary composite structure.

%%%%%%%%%%%%%%%%%%%%%%%%%%%%%%%%%%%%%%%%%
\section{Conclusions}

In this paper, the basic analytic solution for the homogenized effective parameters with complex-valued constituents are examined for acoustic waves. Approximate analytic expressions are developed to describe the regions of the parameter space which exhibit enhanced inertial effects, thereby leading to either an increase in the real or imaginary part of the effective density beyond the bounds of either constituent fluid. The results of this analysis are highlighted using several examples as a function of the filling fraction and are in excellent agreement with Comsol simulations. Realization of a structure which exhibits inertial enhancement is achieved using multi-scale sonic crystals, which allows for two different sonic crystal structures to be used to obtain the desired complex-valued properties of the constituent effective fluids. A multi-scale sonic crystal structure was fabricated using a 3D printer, and acoustic testing was performed using an air-filled impedance tube. The experimental results are in good agreement with the theoretical predictions and demonstrate inertial enhancement.

%%%%%%%%%%%%%%%%%%%%%%%%%%%%%%%%%%%%
\section*{Acknowledgements}
This work was supported by the U.S. Office of Naval Research (Award N000141210216) and by the Spanish \emph{Ministerio de Economia y Competitividad} (MINECO) under contract No. TEC2010-19751.

%\bibliography{EnhancedInertiaBib}

\end{document}